\documentclass[twoside,aps,showpacs,floatfix]{revtex4}
\usepackage{amsmath,amssymb,graphicx,dsfont}
\usepackage[latin1]{inputenc}

\newcommand{\al}{\alpha}
\newcommand{\be}{\beta}
\newcommand{\de}{\mathrm{d}}
\newcommand{\dE}{\Delta}

\newcommand{\s}{\vec{\sigma}}

\newcommand{\atanh}{\text{artanh}}
\newcommand{\eps}{\varepsilon}
\newcommand{\hex}{h_{\mathrm{ext}}}
\newcommand{\Q}{\mathcal{Q}}
\newcommand{\Ph}{\mathcal{P}}

\newcommand{\nn}{\nonumber}
\newcommand{\LL}{\left \langle}
\newcommand{\RR}{\right \rangle}

\newcommand{\beq}{\begin{equation}}
\newcommand{\eeq}{\end{equation}}

\newcommand{\hs }{ \hspace }

\begin{document}
      
\title{ Thermodynamics of the L\'evy spin glass}
\author{K. Janzen$^{1}$, A. Engel$^{1}$ and M. M\'{e}zard${}^{1,2}$}
\affiliation{
${}^1$ Institut f\"ur Physik, Carl-von-Ossietsky Universit\"at, 26111
  Oldenburg, Germany 
\\
${}^2$Laboratoire de Physique Th\'eorique et Mod\`eles Statistiques,
CNRS and Universit\'e Paris-Sud, B\^{a}t 100, 91405 Orsay, France
}

\begin{abstract}
We investigate the L\'evy glass, a mean-field spin glass model with
power-law distributed couplings characterized by a divergent second
moment. By combining extensively many small couplings with a spare 
random backbone of strong bonds the model is intermediate between the
Sherrington-Kirkpatrick and the Viana-Bray model. A truncated version
where couplings smaller than some threshold $\eps$ are neglected can
be studied within the cavity method developed for spin glasses on
locally tree-like random graphs. By performing the limit $\eps\to 0$
in a well-defined way we calculate the thermodynamic functions within
replica symmetry and determine the de Almeida-Thouless line in the
presence of an external magnetic field. Contrary to previous findings
we show that there is no replica-symmetric spin glass phase. Moreover
we determine the leading corrections to the ground-state energy within
one-step replica symmetry breaking. The effects due to the breaking of
replica symmetry appear to be small in accordance with the intuitive
picture that a few strong bonds per spin reduce the degree of
frustration in the system.  

\end{abstract}
\maketitle
\section{Introduction}
Spin glasses have been investigated for about 40 years by now. The 
seminal analysis of Edwards and Anderson \cite{EdAn} revealed that
disorder and frustration are the main ingredients necessary to bring
about the peculiar static and dynamic properties of these
systems. Subsequent analytical and numerical investigations have
indeed shown that model systems with random interactions of the
simplest possible type as, e.g., binary or Gaussian distributions may
qualitatively reproduce various features of experimental spin
glasses \cite{MPV,BiYo}. Trusting in universality and the ubiquitous
efficiency of the central limit theorem no strong dependence of
macroscopic properties on the details of the coupling distribution 
was expected. 

On the other hand many experimental realizations of spin glasses  
involve magnetic impurities placed at random in a non-magnetic
metallic lattice. Mediated by the conduction electrons of the host
material the impurities interact via the RKKY-interaction which
oscillates in sign and falls off with distance $r$ as $1/r^3$. In a
homogeneous sample a given impurity hence interacts with order $r^2$
other impurities a distance $r$ away. Correspondingly this impurity
maintains order $1/J^2$ interactions of strength $J$. Probability
distributions with such power law behavior are markedly different
from simple $\pm J$ or Gaussian distributions. They describe a broad
hierarchy of couplings and do not obey the central limit theorem 
\cite{GnKo}. Well-known representatives are L\'evy distributions
characterized by a power law tail of the form $1/J^{1+\alpha}$ with
the parameter $\al$ ranging between zero and two. Early investigations
of spin glasses with RKKY-interaction \cite{KlBr,Klein} already gave
arguments for a broad distribution of magnetic exchange
fields. More recently spin glass models with a wide hierarchy of
coupling strengths have been used because they can lead to some 
models of finite dimensional spin glasses  which may be well
controlled (in the limit of a very strong hierarchy of couplings)
\cite{NeSt,CMB}.  

The investigation of spin-glass models with power-law distributed
couplings was initiated in 1993 by Cizeau and Bouchaud \cite{CB,Cizeau}
who studied an
infinite-range model using the replica symmetric (RS) cavity
method. The model shows a transition from a paramagnetic
high-temperature phase to a disordered glass phase at a freezing
temperature $T_c$ depending on the parameter $\al$. In their analysis
of the low-temperature phase Cizeau and Bouchaud first provide
arguments for a Gaussian distribution $\Ph(h)$ of local magnetic
fields and then proceed to show that the model exhibits several
unusual properties. The RS entropy becomes negative at sufficiently
low temperature but returns to zero at zero temperature which gave
rise to the speculation that RS may be restored at $T=0$. Moreover,
investigating the local stability of RS they found that the
de~Almeida-Thouless (AT) temperature $T_{AT}$ \cite{AT} is lower than the
freezing temperature $T_c$ suggesting the existence of a finite
temperature interval with a glass phase correctly described by RS. 

The model was re-investigated recently \cite{JHE,Neri,JEM} and it was
found that the distribution of local fields is not a Gaussian and that
the AT-temperature in zero field coincides with the freezing
temperature excluding the possibility of an RS glass phase. In the
present paper we give the detailed derivation of our results
reported in \cite{JEM} and extend them in several directions. We
provide a thorough analysis of the RS properties of the system,
showing that the RS entropy does not vanish when $T\to 0$. We also
characterize the correction which is to be expected from replica
symmetry breaking (RSB) effects by determining the ground state 
energy of the system within one-step RSB. Most of our analysis is done
in the framework of the cavity method, however, we make contact with
the corresponding results from the replica analysis. 

The L\'evy spin glass is intermediate between the two extreme 
prototypes of mean-field spin-glass models, the fully connected
Sherrington-Kirkpatrick (SK) model \cite{SK} and the strongly diluted
Viana-Bray (VB) model \cite{VB,KaSo,MePa86} respectively. Similar to
the SK model 
in the L\'evy glass each spin interacts with all the other spins. The
majority of the couplings is weak (${\mathcal O}(N^{-1/\al})$) as
typical for fully-connected models. On the other hand due to the heavy
tails of the coupling distribution on top of this background of weak
couplings there is a backbone built from a few (${\mathcal O}(1)$)
strong couplings per spin which remain ${\mathcal   O}(1)$ for
$N\to\infty$ similar to the VB model. A somewhat related situation is
given by composite systems \cite{RaSa} for which the properties of the
weak and the strong bonds are defined separately. 

The decisive question is which
properties of the L\'evy glass are exclusively determined by the 
strong bonds and which also feel the influence of the many weak
ones. To elucidate this point we will often consider what we call the
{\it truncated} model in which all couplings weaker than a certain
threshold $\eps$ are neglected. This technique has been crucial in the
 recent developments on the L\'evy spin glass problem,  see
 \cite{Neri,JEM}. We are then dealing with a spin-glass on an
 Erd\"os-R\'enyi random graph and use techniques developed for the
 cavity analysis of these systems
 \cite{MezParBethe,MezMonBook}. Eventually, we have to investigate the
 crucial limit $\eps\to 0$ to recover the original L\'evy glass.

The paper is organized as follows. In section II we introduce the
model and the basic notation. Section III is devoted to the
determination of the freezing temperature $T_c(\al)$ and the
RS distribution of local fields. In section IV we derive
expressions for the thermodynamic functions like the free energy, the
internal energy and the entropy and discuss the RS phase diagram of
the L\'evy glass. Section V contains the determination of the AT-line
of the L\'evy glass. In section VI we calculate the corrections to the
ground state energy resulting from one-step RSB. Finally, section VII
contains some conclusions. 

\section{The model}
\label{model}
We consider a system of $N$ Ising spins $S_i=\pm 1,\, i=1,...,N$ with
Hamiltonian 
\begin{equation}
  \label{eq:H}
  H(\{S_i\})=-\frac{1} {2} \sum_{(i,j)} J_{ij} S_i S_j- \hex \sum_{i}
  S_i  \; , 
\end{equation}
where the sum is over all pairs of spins, and $\hex$ denotes an
external magnetic field. 
The couplings \mbox{$J_{ij}=J_{ji}$} are independent, identically
distributed random variables drawn from the distribution 

\begin{equation}
  \label{eq:defP}
 P_{\al, N}(J)=\frac{\al}{2
 N}\frac{1}{|J|^{\al+1}}\;\theta(|J|-N^{-1/\al})\; ,  
\end{equation}

where $\theta$ denotes the Heaviside function and the scaling of 
the couplings with $N$ ensures that the free energy of the system is
extensive. The most prominent feature of the coupling distribution
(\ref{eq:defP}) is its power-law tail for large values of $|J|$. In
fact we will see that all macroscopic properties of the system depend
only on the overall scale of couplings and the exponent $\al$
characterizing these tails. This implies in particular that all our
result apply also to a spin glass with coupling distribution given by
a symmetric L\'evy distribution $P^L_\al$ defined by the characteristic
function   
\begin{equation}\label{eq:deffullLevyP}
\widehat{ P_{\al}^L}(q)= \int d J \, P^L_\al(J)\, e^{iqJ}=
  \exp \Big( - \frac{  \tilde J_{1,\al} }{N}|q|^\al  \Big)\, , 
\eeq
with
\beq
 {\tilde J_{1,\al}} =\frac{\al \pi}{2 \sin(\pi  \al /2 )
  \Gamma(\al+1)}\, .  \label{eq:deffullLevyP2}
\end{equation}
In the large $N$ limit the expectation values of any well-behaved
function of $J$ taken with respect to the distribution (\ref{eq:defP})
and (\ref{eq:deffullLevyP}) respectively coincide.  

In the present paper we will
assume that \mbox{$\al \in ]1,2[$ } implying a finite average of
$|J|$. With the Hamiltonian (\ref{eq:H}) being linear in the 
$J_{ij}$ we expect that for these values of $\al$ the 
thermodynamic potentials will be self-averaging. 


\section{Distribution of local fields}
\subsection{Self-consistent equation for the local field distribution}
The central quantity in the replica symmetric cavity analysis of spin
glasses is the distribution $\Ph(h)$ of local fields $h_i$ that
parametrize the marginal thermal distribution of spin $S_i$ at site
$i$ \cite{MPV,MezMonBook}. Adding to a system of $N$ spins
$S_i,\,i=1,...,N$ another spin $S_0$ with couplings
$J_{0i},\,i=1,...,N$ one finds for the local field at
the new site the equation \cite{MezParBethe,CB}
\begin{eqnarray}\label{eq:up}
 h_0&=&\hex+\sum_{i=1}^N  u(h_i,J_{0i}) \, ,
\end{eqnarray}
where 
\begin{equation}
  u(h,J)=\frac{1}{\beta}\atanh\big(\tanh(\beta h) \tanh(\beta
  J)\big)\, .
\end{equation}
The new field $h_0$ is a random quantity due to the randomness in the $h_i$
and $J_{0i}$. The update equation (\ref{eq:up}) may hence be turned
into a self-consistency condition by averaging over the distribution
of $h_i$ and $J_{0i}$ and requiring that the statistical properties at
site $i=0$ should be equivalent to those at all other sites. Accordingly
\begin{eqnarray}\nonumber
\Ph(h) &=& \int \!\!\prod_i \de h_i \, \Ph (h_i)\!\!\int\!\! \prod_i
  \de J_{0i}\, P_{\al,N}(J_{0i})\; 
  \delta\big(h-\hex-\sum_{i=1}^N u(h_i,J_{0i})\big)\\ \nonumber
  &=& \int\frac{\de s}{2\pi} \exp \left( is(h-\hex)\right)
    \left[\int \!\!\de h' \,\Ph (h')\!\!\int\!\!\de J\, P_{\al,N}(J)\; 
          \exp\big(-is\, u(h',J)\big) \right]^N\nonumber\\\nonumber
  &=&  \int\frac{\de s}{2\pi}  \exp \left( is(h-\hex)
  +N \ln   \left[1+
    \frac{\al}{ 2 N}  \int \de h'\, \Ph (h') \int \frac{\de
  J}{|J|^{\al+1}} \big( \cos( s\, u(h',J)) -1 \big) 
  \theta\left(|J|-N^{-\frac{1}{\al}}\right)
 \right]
 \right)
  \\
     &{ N \to \infty} \atop \to& \int\frac{\de s}{2\pi} \exp{\Big(is(h-\hex)+ 
    \frac{\al}{2}  \int \de h' \, \Ph (h')  
    \int \frac{\de J}{|J|^{\al+1}}
          \big(\cos( s\, u(h',J))-1\big)\Big) }  \, .
\label{eq:sc-eq} 
\end{eqnarray}
In the second line we used the statistical independence of the
distributions at different sites. In the third one we inserted the
explicit form (\ref{eq:defP}) of $P_{\al,N}(J)$ and took advantage of
the fact that it is an even function of $J$. Note that splitting off
the leading $1$ in the square brackets makes the $J$-integral well
defined for $N\to \infty$ since $\cos(u(J,h))-1$ is quadratic in $J$
for small $J$ and therefore suppresses the potential divergence
of the integral for $|J|\to 0$. This allows us to safely perform the
limit $N\to\infty$ in the last line. 

Eq.(\ref{eq:sc-eq}) is the central equation of the RS cavity
approach. From its solution one gets $\Ph(h)$ which in turn determines all
equilibrium properties of the system within replica symmetry. The same
equation was also obtained in \cite{JHE} by using the replica method. 

\begin{figure}
  \begin{center}
    \includegraphics[height=0.35\textwidth]{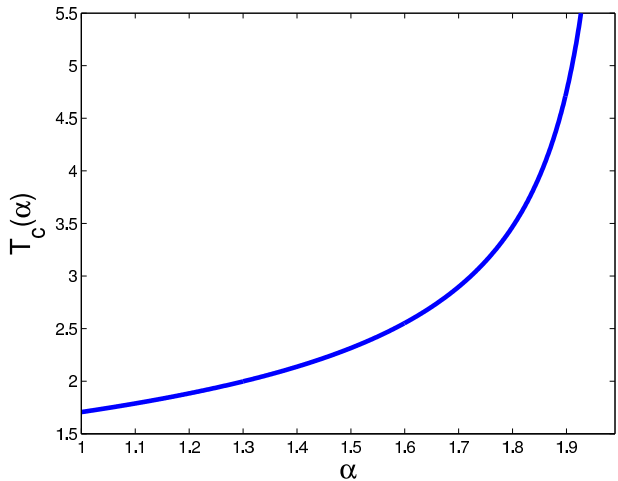}
  \end{center}
  \caption
      {
	Spin glass transition temperature $T_c(\al)$ of an infinite-range 
	spin-glass with couplings drawn from the distribution $P_{\al,N}$
	defined in (\ref{eq:defP}).
      } 
      \label{fig:Tc}
\end{figure}

For high temperatures and zero external field we must have 
$\langle S_i \rangle=0$ for all $i$ and hence all local fields must
vanish. Indeed $\Ph (h) = \delta (h)$ is a solution of
(\ref{eq:sc-eq}) for all temperatures. To test its stability one
starts with a distribution $\Ph(h)$ with a small second moment and
investigates whether it grows or shrinks under iteration. In this way  
one finds that the paramagnetic solution $\Ph (h) = \delta (h)$
becomes unstable below the critical temperature 
\begin{equation}\label{eq:Tc}
T_c(\alpha)=\left[\al \int_0^\infty \frac{\de x}{x^{\al+1}} 
   \tanh^2 x\right]^{\frac{1}{\al}} \, .
\end{equation}
This result was already obtained in \cite{CB} and \cite{JHE}. The
dependence of $T_c$ on $\al$ is shown in Fig.~\ref{fig:Tc}. 
Note that $T_c$ diverges as $\al \to 2$. On the other hand it is known
that the L\'evy distribution tends to a Gaussian in this limit and we
would hence expect to reproduce the results for the SK-model for $\al
\to 2$. This is indeed the case, however, the L\'evy distribution
(\ref{eq:deffullLevyP}) approaches a Gaussian with {\it divergent}
variance, $\tilde J_{1,\al=2}$,
cf. (\ref{eq:deffullLevyP2}). Rescaling $T_c$ with  $\tilde J_{1,\al}$
we indeed recover the classical results for the SK-model \cite{SK}. 

Below the freezing temperature $T_c(\al)$ no analytical solution
of (\ref{eq:sc-eq}) is available. We note in particular that a
Gaussian ansatz for $\Ph (h)$ as advocated in \cite{CB} does not
reduce (\ref{eq:sc-eq}) to a self-consistent equation for the 
variance: Plugging in a Gaussian at the r.h.s. of (\ref{eq:sc-eq})
does not produce a Gaussian at the l.h.s. The only way to determine
$\Ph (h)$ in the spin-glass phase $T<T_c(\al)$ is hence by numerical
methods. 

\subsection{Numerical determination of the local field distribution} 
\label{sec:SolSC}
We have implemented two ways to numerically solve
(\ref{eq:sc-eq}). The first consists in a straight iteration of the
equation. From the $n$-th approximation  $\Ph^{(n)}(h)$ of the unknown
distribution $\Ph(h)$ we determine 
\beq\label{eq:iter1}
Q^{(n)}(s)=\exp\Big( \frac{\al}{2} \int \de h' \, \Ph^{(n)} (h')  
    \int \frac{\de J}{|J|^{\al+1}} \big(\cos( s\, u(h',J))-1\big)\Big) 
\eeq
by numerical integration. The next approximation, $\Ph^{(n+1)}(h)$, is
then obtained via Fourier transform 
\beq\label{eq:iter2}
 \Ph^{(n+1)}(h)=\int\frac{\de s}{2\pi}\; e^{is(h-\hex)}\,Q^{(n)}(s) \, . 
\eeq
Starting initially from a uniform or normal distribution $\Ph^{(0)}(h)$
the procedure converges rather quickly. In order to save computation
time we subdivide the $J$-integral in (\ref{eq:iter1}) and approximate
the small and large $J$ part by analytical expressions. Moreover we
use the {\it Fast Fourier Transform} with 256 or 512 nodes
to perform the second step (\ref{eq:iter2}). Since in general results
for several values of $\be$ are needed it is convenient to use the
final result for one $\beta$ as the initial distribution in the
iteration for the next one. In this way smooth and
accurate approximations for $\Ph(h)$ can be obtained in reasonable
time. Moreover, in the limit $\beta\to\infty$ the integral in
(\ref{eq:iter1}) may be simplified which makes the method very
efficient for determining $\Ph(h)$ at zero temperature. 

Alternatively population dynamics as introduced for diluted spin
glasses in \cite{MezParBethe} may be used to solve
(\ref{eq:sc-eq}). Compared to the direct iteration discussed above
this method has two disadvantages: Firstly, it is statistical in
nature and therefore one has to cope with intrinsic
fluctuations. Secondly its application to the L\'evy glass needs the
introduction of an additional cut-off parameter $\eps$ for the
coupling strength. On the other hand population dynamics has a big
advantage which outweighs the above mentioned drawbacks: In a
generalized form it may also be used for the investigation of the RSB 
phase (cf. section \ref{sec:1RSB}). 

The most direct way to map the L\'evy glass onto a diluted spin
glass amenable to population dynamics is by using the truncated model,
i.e. by simply neglecting all bonds with modulus less than some
threshold $\eps$. The number $(K+1)$ of remaining bonds per site is
then a Poissonian random variable with mean $\eps^{-\al}$. The
distribution of the remaining bonds is given by   
\beq
P_{\al, s}(J)=\frac{\al\,\eps^\al}{2 |J|^{\al+1}}\, 
    \theta(|J|-\eps)\, . 
\label{eq:Pstrong}
\eeq
Population dynamics may now be applied without further ado:
Choosing $K$ from its Poisson distribution and selecting at random $(K+1)$
values $h_k$ from an initial seed one replaces $h_{K+1}$ with  
\beq
 h_{\mathrm{new}}=\hex+\sum_{k=1}^{K}  u(h_k,J_k) 
 \label{eq:popdyn1}
\eeq
until the histogram of the $h_k$ no longer changes significantly. This
procedure has to be performed for successively smaller values of
$\eps$ from which the asymptotic result for $\eps\to 0$ may be
extracted. 

Although this method works in principle, its convergence for $\eps\to
0$  is slow.  We found a significant speed-up of the
algorithm by using the following modification. Instead of neglecting
the weak bonds altogether we subsume them into a Gaussian random
variable $z$ with zero mean and a variance determined
self-consistently. From the distribution 
\beq 
P_{\al, w}(J) = 
 \frac{1}{N-\eps^{-\al}}
 \frac{\al}{2 |J|^{\al+1}}\,
 \theta\big(|J| - N^{-1/\al}\big)\,
 \theta(\eps -|J|)
\eeq
of weak bonds we find for this variance 
\begin{eqnarray} \nonumber
 \overline{z^2}
&=&
 \overline{ \left(\sum_{i=K+1}^{N} u(h_i,J_{0i})\right)^2}
=(N- \eps^{-\al} )
\int \de h  \,\Ph (h) 
\int \de J \,  P_{\al, w}(J) \, u^2(h,J) 
\\
&N\to \infty \atop \to& 
 \int \de h  \, \Ph (h)
\int_0^\eps
   \frac{ \al \,\de J}{J^{\al+1}} \;u^2(h,J) \, = :
\LL
\int_0^\eps
   \frac{ \al \,\de J}{J^{\al+1}} \;u^2(h,J) \,
\RR_h  \,.
\label{eq:varsmall}
\end{eqnarray}
Here as in the rest of the paper $ \LL \, \cdot \, \RR_h $ denotes the
average with respect to the distribution of the local fields $\Ph(h)$
and the over-bar indicates the quenched average over the appropriate
distribution of bonds. 

By combining the update (\ref{eq:varsmall}) for $\overline{z^2}$ with
the {\it noisy} population dynamics algorithm
\begin{eqnarray}
 h_{\mathrm{new}}=&=&\hex+\sum_{k=1}^{K}  u(h_k,J_k) + z
 \label{hlargesmall}
\end{eqnarray}
we treat the dominant contributions from the strong couplings exactly,
and include the influence of the weak couplings in an approximate
way. This modification of (\ref{eq:popdyn1}) has at least two
advantages. Most importantly its numerical implementation showed that
the final extrapolation to $\eps\to 0$ is much smoother.  
Moreover, in the opposite limit, $\eps\to\infty$, all couplings are
included in the Gaussian variable $z$ and we may check the algorithm 
by comparison with the results obtained in \cite{CB}. 

The integrals entering (\ref{eq:varsmall}) are still computer demanding. 
To save some computer time it is useful to tabulate the function
\begin{eqnarray} \nonumber
h \mapsto \int_0^\eps \frac{ \al \,\de J}{J^{\al+1}} \;u^2(h,J)
\end{eqnarray}
for an appropriate interval of values of the local field $h$ before the 
update procedure. The determination of the variance
(\ref{eq:varsmall}) is then reduced to a simple integral at each
update. 

\begin{figure}
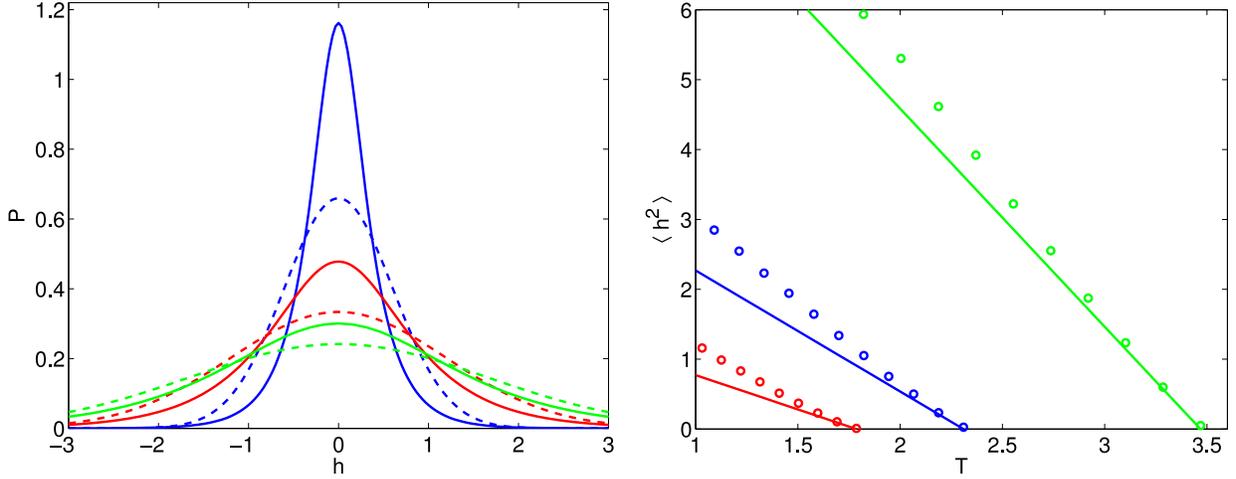

 \includegraphics[height=0.35\textwidth]{Pofh.epsi}
 \hspace{0.2cm}
 \includegraphics[height=0.35\textwidth]{h2c.epsi}
 \caption{
 Left: $\Ph (h)$ for $\al = 1.1,\; T= 0.9\,T _c,\; 0.6\,T_c,\;
 0.1\,T_c$ (center from top to bottom). Full line: results from the
 iterative solution of the self-consistent equation (\ref{eq:sc-eq}).
 Dotted line:  Gaussian approximation from \cite{CB}. Right: Second
 moment of $\Ph (h)$ for $\al=1.1,1.5,1.8$ (from left to right)
 as a function of temperature. Symbols: As
 determined from  the iteration results for $\Ph (h)$. Lines: From the
 analytical expansion of (\ref{eq:sc-eq}) around $T_c$,
 cf. (\ref{eq:h2expan}).}  \label{fig:P(h)} 
\end{figure}

The left part of Fig.~\ref{fig:P(h)} shows the distribution of local
fields $\Ph(h)$ in zero external field as obtained by direct iteration
of eq.~(\ref{eq:sc-eq}). We find practically indistinguishable results
by running $10^3$ iteration of population dynamics with $\eps\leq 0.3$ 
using $10^4$ members in the population. Markedly different,
however, are the Gaussian distributions proposed in 
\cite{CB} which are also shown in Fig.~\ref{fig:P(h)}. The reason for
these differences lies in the fact that that the sum in (\ref{eq:up})
is dominated by a few large contributions. Consequently, although the
different terms in the sum are statistically independent and all have
finite second moments, the central limit theorem may not be invoked 
since the Lindeberg criterion is not fulfilled.

To further check our numerical results we have determined the second
moment of $\Ph(h)$ as a function of temperature close to 
$T_c(\al)$, cf. the right part of Fig.~\ref{fig:P(h)}. The variance
tends to zero when the temperature approaches $T_c$ from below as it
should. Moreover, the slope coincides with the one following from the
analytical expansion of (\ref{eq:sc-eq}) for small $T-T_c$ which gives  
\begin{eqnarray} \label{eq:h2expan}
 \LL h^2 \RR_h  = - \frac{\al}{ 2} T_c(\al) \, \big (T-T_c(\al) \big )
 +{\cal O}\big( (T-T_c(\al))^2\big )\; .
\end{eqnarray}

Our solutions for $\Ph(h)$ are similar to those given in Fig.~2 of
\cite{Neri}. However, our results for the Gaussian approximation are
significantly different from those shown there. 

\section{Replica Symmetric Thermodynamics}\label{Sec:RS}
In this section we give expressions for the thermodynamic
potentials - free energy, internal energy and entropy - which, on the
replica symmetric level, are all functionals of the local field
distribution $\Ph (h)$ determined above. 
We first derive a differential equation for the free energy per spin 
$f(\beta,\hex)$ of the model defined by eqs.~(\ref{eq:H}) and
(\ref{eq:defP}) within the cavity method. Next we consider the
truncated model and derive an expression for the free energy and the
internal energy using the cavity method for diluted spin glasses. We
then show that, in the limit $\eps\to 0$, this free energy fulfills
the differential equation derived before. We also show how  the
same
expressions for the thermodynamic functions can be derived from the replica
method. Finally we discuss the most salient features of the RS
thermodynamics of the L\'evy spin glass.  
\subsection{Differential equation for the free energy per spin}
\label{sec:TDode}
Assuming that the free energy per spin $f(\beta,\hex)$ is
self-averaging in the thermodynamic limit it can be related to the
shift in free energy due to the addition of one spin via 
\begin{eqnarray} 
 \beta f(\beta,\hex) =  - \lim_{N\to \infty }  \frac{ \overline{\log
   Z_N(\be,\hex)}}{N} = - \lim_{N\to \infty } \Big(\, \overline{ \log
   Z_{N+1}(\be,\hex)} - \overline{ \log  Z_{N}(\be,\hex) \,} \Big)\, ,
\label{eq:betaf}
\end{eqnarray}
where the disorder average is taken with respect to the coupling
distributions $P_{\al,\, N+1}$ and $P_{\al,\, N}$ respectively. 
Due to the explicit dependence of the coupling distribution on the
number of spins the comparison between systems of different size needs
some care \cite{MPV}. In the present case the slight change of the
coupling distribution (\ref{eq:defP}) when going from $N$ to
$(N+1)$ spins is absorbed in the rescaled parameters  
\begin{eqnarray} \nn
\be' = \be \left (\frac{N+1}{N}  \right) ^{\frac{1}{\al}} 
\qquad
\mathrm{and}
\qquad
\hex'=\hex\left (\frac{N+1}{N}  \right) ^{-\frac{1}{\al}}\, .
\end{eqnarray}
Splitting off the terms depending on the new spin $S_0$ we find 
\beq\label{eq:hZZ} 
\log   \frac{Z_{N+1}(\be ', \hex')}{Z_N(\be, \hex )}
= \log  \sum_{S_0}\sum_{ \{S_i\}_{ i=1..N}}  P(\{S_i\}) 
    \exp\big( \be S_0 \sum_{i=1}^N J_{0i} S_i + \be \hex S_0  \big)\; .
\eeq
The clustering property of the pure state in the absence of the new spin,
i.e. the statistical independence of all spins interacting with the
newcomer implies  
\begin{eqnarray}
 P(\{S_i\})
= \prod_{i=1}^N \frac{ \exp(\beta h_i S_i)}{2 \,\cosh (\be h_i)}\, ,  
\label{eq:clustering}
\end{eqnarray}
which when used in (\ref{eq:hZZ}) yields 
\begin{eqnarray} \nonumber
\log   \frac{Z_{N+1}(\be ', \hex')}{Z_N(\be, \hex )}
&=& \sum_{i=1}^N \log \cosh(\be J_{0i})
  + \frac{1}{2} \sum_{i=1}^N 
    \log \big( 1- \tanh^2(\be h_i) \tanh^2(\be J_{0i} ) \big)
  + \log \big( 2 \cosh( \be h_0 ) \big) .
\label{eq:ZZ}
\end{eqnarray}
Averaging (\ref{eq:ZZ}) over the disorder then gives  
\begin{eqnarray} \nn
\overline {\log   \frac{Z_{N+1}(\be ', \hex')}{Z_N(\be, \hex )}}
&=& N  \int \de J \, P_{\al,N}(J) \log  \cosh (\be J) 
+\LL 
 \log \big( 2 \cosh ( \be h )\big)                        
\RR_h\\
&&+\frac{N}{2} \int \de J \, P_{\al,N}(J) 
\LL \log \big(1-\tanh^2(\be h) \tanh^2(\be J) \big)
\RR_h \, .
\label{eq:quenchedZZ}
\end{eqnarray}
The limit $N\to\infty$ may now be taken on both sides of this
equation. Using (\ref{eq:betaf}) and neglecting $\mathcal{O} (1/N)$
contributions we find for the l.h.s.
\begin{equation}\nn
\overline{ \log  \frac{Z_{N+1} \left(\be',\hex' \right) }{Z_N (\be,
    \hex)} } = - \be \, f - \frac{ \be }{ \al }  \frac{\partial \,
    \big( \beta  \,f \big)} {\partial \beta} \,  + \frac{ \hex }{ \al
    } \frac{\partial \,  \big( \beta  \,f \big)} {\partial \hex} =
     -\be f - \frac{ \be }{ \al } \big ( e +  \hex m \big )\, ,
\end{equation}
where we have used the thermodynamic relations for the internal energy
and magnetization per spin respectively
\begin{eqnarray} \nn
 e =\frac{\partial \,  \big( \beta  \,f \big)} {\partial \beta} \
 \qquad 
\be \,m= -\frac{\partial \,\big( \beta\,f \big)} {\partial \hex} \, . 
\end{eqnarray}
Combining this result with the limit of the r.h.s. gives rise to the
differential equation 
\begin{eqnarray} 
\hspace{-0.4cm}
\be \, f + \frac{ \be }{ \al } \big ( e +  \hex m \big )
\hspace{-0.1cm}
&=&
\hspace{-0.1cm}
-  \Big \langle \log \big ( 2 \cosh( \be h) \big) \Big \rangle_h 
\hspace{-0.15cm}
 -
\hspace{-0.1cm} \int 
\hspace{-0.1cm} \frac {  \al \, \de  J }{ 2 |J| ^{\al+1}} 
\left [\log \cosh ( \be J)    +   \frac{1}{2}   \Big \langle
\log  \big(1-\tanh^2(\be h) \tanh^2(\be J) \big )\Big \rangle_h
\right] .
\label{eq:fem}
\end{eqnarray}
For zero external field this equation was already derived in
\cite{Cizeau} where, exploiting the assumption of a Gaussian
distribution of local fields, also an explicit solution was
constructed. However, in view of the fact that the correct 
distribution of local fields is non-Gaussian and is not available
analytically, a straight integration of (\ref{eq:fem}) to find
$f(\beta,\hex)$ is difficult. Instead we will use two different
ways to derive an expression for the free energy per spin and verify
that it indeed fulfills (\ref{eq:fem}).

\subsection{Free energy of the truncated model}
\label{sec:TDtrunc}
When neglecting all bonds with a strength less than a threshold $\eps$
the L\'evy spin glass is converted into a spin glass on a locally
tree-like random graph $\mathcal{G}_{N,\eps^{-\al}}$ with $N$ sites, mean
connectivity $\eps^{-\al}$, and a coupling distribution given by 
(\ref{eq:Pstrong}). 
We may therefore use methods from the cavity analysis of the Bethe
spin glass \cite{MezParBethe} with only minor changes due to the
fluctuating connectivity in our model.
The free energy per spin of the truncated model is given in terms of
two different free energy shifts according to 
\begin{eqnarray}
\be f_\eps(\hex,\be ) = -\frac{1}{2}
 \lim_{N\to \infty} \overline{ \big( \log Z_{N+2}- \log Z_N  \big) }
=\frac{1}{2}  \left(  \overline{(K+1) \Delta F^{(2)}_\eps}-
 \overline{2 K \Delta F^{(1)}_\eps} \right). 
\label{eq:free_diluted}
\end{eqnarray}
Here $ \Delta F^{(1)}_\eps$  corresponds to the free energy shift due
to the addition of a single spin and $ \Delta F^{(2)}_\eps$ to that
due to the addition of two spins connected by a bond. 
An intuitive explanation for this relation can be obtained by considering 
two operations acting on the graph $\mathcal{G}_{N, \eps^{-\al}}$ of the truncated model.
Removing $2K$ vertices from this graph leads to a cavity graph 
where some spins lack neighbors,
here as before \mbox{$(K+1)$} is a  poissonian with mean $\eps^{-\al}$.
Adding \mbox{$(K+1)$} new pairs of spins $\sigma_0, \tau_0$ 
connected by a bond $J_0$ to the system, and connecting them
to the free spins produced by the first operation leads to 
a  $\mathcal{G}_{N+2, \eps^{-\al}}$ graph where the connectivity
remains unchanged whereas the number of vertices is increased by $2$.
The resulting expression for the free energy per spin reads :
\begin{eqnarray}  \nn
\be f_\eps  &=&
 - \frac{ \overline{(K+1)} }{2} \int  \de J \, P_{\al,s } (J) \, \log
 \cosh \be J 
 + \frac{ \overline{K}     }{2} \int  \de J \, P_{\al,s } (J) \, \LL
 \log  \big (1-\tanh^2( \be J)  \tanh^2( \be h)  \big) \RR_h \\ \nn 
&&-\LL \log 2 \cosh \be h \RR_h 
  -\frac{ \overline{(K-1)} }{4} \int  \de J \, P_{\al,s } (J) 
\LL \log \big (1-\tanh^2( \be J)  \tanh^2( \be h) \tanh^2(\be h')
\big)  \RR_{h, h'}\, . 
\end{eqnarray}
Taking into account the $\eps$-dependence of the distributions of the
connectivity and the couplings strength the limit $\eps \to 0 $ may be
performed and we obtain for the free energy per spin of the original
model 
\begin{eqnarray}  \nn
\be f  &=&  -
\int \frac{ \al \, \de J}{4\, |J|^{\al +1}} 
\Big[
  \log \cosh(\be J)
  +
  \LL
  \log \big (1-\tanh^2( \be J)  \tanh^2( \be h)  \big) 
  \RR_{h}
  \Big]-\LL \log \big( 2  \cosh (\be h ) \big) \RR_h \\
\label{eq:f_final} 
 &&
+ \int \frac{\al \, \de J}{ 8\, |J|^{\al +1}}
  \LL \log \big (1-\tanh^2( \be J)  \tanh^2( \be h) \tanh^2(\be h')  \big) 
  \RR_{h,h'}.
\end{eqnarray}
We now turn to the determination of the internal energy $e_{\eps}$ per spin. 
There are two contributions: one due to the interactions and one due
to the external field, $e_\eps=e_\eps^{\mathrm{link}} - \hex m$. 
To obtain the link contribution we follow the steps in \cite{MezParBethe} 
and add a coupling $J_{ij}$ to the system. 
For the L\'evy case it is convenient to rewrite the expression for the 
energy of this link  obtained in \cite{MezParBethe}  as 
\begin{eqnarray} \nn
 E_{ij}= -J_{ij} \LL S_i S_j \RR
=  -\left .\frac{\partial}{ \partial \be '} 
     \log \Big(
     \cosh( \be' J_{ij}  ) 
     \big(1+\tanh(\be' J_{ij} ) \tanh (\be h^{}_i) \tanh( \be h^{}_j )  \big)
      \Big) 
 \right \vert_{\be = \be '}\, ,
\end{eqnarray}
here $h_j$ and $h_j$ denote the local fields in the absence of the new
link at the site $i$ and $j$ respectively.
A link connects two spins, each of which interacts on average with
$\eps^{-\al} = \overline{K+1}$ neighbors. 
After the average over the quenched disorder the system is homogeneous.
The link contribution to the internal energy is hence related to
the average of $E_{ij}$ by 
\begin{eqnarray} \nn
  e_\eps^{ \mathrm{link}}
  &=&
  \overline{\frac{K+1}{2}   E_{ij}}
  =
  -\frac{\eps^{-\al}}{2}  \int \de J \, P_{\al, s}(J)\left .\LL
  \frac{\partial}{ \partial \be '} \log   \Big( \cosh (\be' J)  \big
  (1+\tanh(\be' J ) \tanh (\be h) \tanh( \be h' ) \big ) \Big )
  \RR_{h,h'}  \right \vert_{\be = \be '}\\ \nn 
  &=&  -\frac{1}{2}  \int_{|J|> \eps}  \frac {    \al \, \de  J
  }{ 2 |J| ^{\al+1}} \left . \frac{\partial}{ \partial \be '} \left[
    \log    \cosh (\be' J)  + \frac{1}{2} \LL \log \big (1-\tanh^2(\be'
    J ) \tanh^2 (\be h) \tanh^2( \be h' ) \big ) \RR_{h,h'} \right]
  \right \vert_{\be = \be '} \, , 
\end{eqnarray}
where we skipped all contributions which vanish due to the symmetry of the
$P_{\al, s}$ distribution. 
Using in addition \mbox{$m=\LL \tanh(\be h)\RR_h$} we find  
for the internal energy of our original model : 
\begin{eqnarray} 
  \nn
  e &=& \lim_{\eps \to 0}  e_\eps^{ \mathrm{link}}  - \hex \LL
  \tanh(\be h)\RR_h\\ \nn 
  & =&- \left. \frac{\partial}{ \partial \be '}
  \be'^\al \right  \vert_{\be = \be '} \frac{1}{2} \int
  \frac {  \al \, \de  J }{ 2 |J| ^{\al+1}} \left[   \log    \cosh ( J)   
    + \frac{1}{2} \LL \log \big (1-\tanh^2( J ) \tanh^2 (\be h) \tanh^2(
    \be h' ) \big ) \RR_{h,h'}\right] - \hex \LL \tanh(\be h)\RR_h\\ 
  &=&- \frac{\al}{ 2 \be} \int  \frac {   \al \, \de  J }{
    2 |J| ^{\al+1}} \left[   \log    \cosh ( \be J)  + \frac{1}{2} \LL
  \log \big (1-\tanh^2(\be J ) \tanh^2 (\be h) \tanh^2( \be h' ) \big
  )  \RR_{h,h'}\right] - \hex \LL \tanh(\be h)\RR_h \, .
\label{eq:e_final}
\end{eqnarray}
The results obtained for $f$, $e$  and $m$ fulfill the differential
equation (\ref{eq:fem}) derived in \ref{sec:TDode}. 
The replica symmetric
thermodynamics of the truncated model is hence in the limit $\eps\to
0$ equivalent to that of the L\'evy spin glass. Note that the above
reasoning relies on the fact that the limits $N\to\infty$ and $\eps\to
0$ commute. However, since all expressions depend smoothly on the
cutoff parameter for $\eps \to 0$ we believe that this is indeed the
case. In order to further substantiate our results (\ref{eq:f_final})
and (\ref{eq:e_final}) we re-derive them in the next subsection using
the replica approach. 

\subsection{Free energy from the replica approach }\label{sec:TDrepl}
Within the replica approach the free energy per spin is related to the
replicated partition function $\overline{ Z^n}$ via 
\begin{eqnarray}
 \be f 
 = -\lim _{n \to 0} \lim_{N \to  \infty} \frac{1}{N n }  \log \overline{ Z^n}. 
\end{eqnarray}
Due to the slow decay of $P_{\al, N}(J)$ the quenched average of
$\overline{Z^n}$ does not exist for $n\neq 0$. 
As in \cite{JHE} we therefore use imaginary temperatures 
$  \be = -ik \;, k \in \mathbb{R}$ at intermediate steps of 
the calculation, i.e. before the limit $n \to 0$ is taken.

For integer values of $n$, the quenched average of the partition
function reads 
\begin{eqnarray}\label{repZ1}
\overline{ Z^{n}(-ik)}&=& \sum_{ \{ S^{a}_i \}}\int
 \prod_{i<j}\de J_{ij} \, P_{\al , N} (J_{ij})  \, \exp (-ik J \vec S_i
 \cdot \vec S_j) \\ \nn  
&=&  
 \sum_{ \{ S^{a}_i \}}
 \exp \left( \frac{1}{2} \sum_{(i,j)}
 \log \left (1+ \frac{1}{N} \int  \frac{ \al  \,\de J}{ 2 \,|J|^{\al+1}}
 \left[ \cos \left( \,k\,J \vec S_i \cdot \vec S_j  \right )  -1
 \right ]\right) + \mathcal{O}(N^{\gamma}) 
 \right )  , 
\end{eqnarray}
where $\gamma <1$ for all values of the parameter $\al$ considered
here. 
Eq. (\ref{repZ1}) can now be transformed into a $2^n$-dimensional integral 
over order parameters
\cite{Remi}  
\begin{eqnarray}
c(\s) =\frac{1}{N} \sum_{i=1}^N \delta(\s, \vec S_i)
\label{eq:repOP}
\end{eqnarray}
where $\s = \{\sigma_a\}$ denotes an $n$-component Ising vector and
$\delta$ is the Kronecker-$\delta$.
The partition function acquires the form 
\begin{eqnarray}\label{repZ}
\overline{ Z^{n}}
= \int \prod _{ \s} \de c \left(  \s \right)  
  \delta \left ( \sum_{ \s} c \left(  \s \right)-1 \right)
  \exp   \Big (-N (-ik)  f_{trial} \big ( \left \{c \left  (\s
    \right)\right\} \big)\Big)\; , 
\end{eqnarray}
and is  evaluated by the saddle-point method for $N\to\infty$.
The free energy per spin is then determined by 
\beq\label{eq:rsf}
 f  =  \lim _{n \to 0} \frac{1}{ n }  f_{trial} \big(\{c_0 ( \s )\}
   \big) 
\eeq
where $c_0$ minimizes the trial free energy 
\begin{eqnarray}
(-ik)   
f_{trial}\big( \{c (\s )\} \big) = \sum_{\s }  c( \s ) \,  \log  c( \s)
-\frac{ 1 }{2}\sum_{ \s ,\s ' }  c( \s )\,  c(\s ' ) \,
\int \frac{\al \, \de J}{ 2 |J|^{\al +1}}
 \left [  \cos \left( \,  k J \,  \s \cdot  \s ' \right)   -1 \right].
\label{eq:freetrial}
\end{eqnarray}
The corresponding saddle-point equation reads
\begin{eqnarray}
0= 1+\Lambda(n)+ \log c_0( \s)+
  \sum_{\s '} c_0(\s ')
\int \frac{\al \, \de J}{ 2 |J|^{\al +1}}
 \left [  \cos \left( \,  k J \,  \s \cdot  \s ' \right)   -1 \right]
\label{eq:seq_rep}
\end{eqnarray}
where the Lagrange multiplier $\Lambda(n)$ accounts for the 
constraint $\sum _{\s } c(\s ) =1 $ resulting from (\ref{eq:repOP}).
Within the replica symmetric assumption  $c_0(\s)$ depends on the 
sum $\sum_{a=1}^n \sigma_a$ only, and is related to the distribution
$\Ph(h)$ of local fields via \cite{Remi}
\begin{eqnarray}\label{eq:RSrepl}
c_0^{RS}(\s) = \int \de h \, \Ph (h) 
\frac{ \exp(
-ik h \sum_{a =1}^n \sigma_a)}
     {\big( 2 \cosh 
(-ik h) \big )^n       }.
\end{eqnarray}
The saddle-point equation (\ref{eq:seq_rep}) for $c_0(\s)$ can then be
transformed into a self-consistent equation for $\Ph (h)$ which
coincides with (\ref{eq:sc-eq}). 
Using the RS ansatz (\ref{eq:RSrepl}) the sums in (\ref{eq:freetrial})
can be performed leading to expressions for which the $n \to 0 $ limit
can be taken (for details see \cite{JHE}). 
The value of the Lagrange multiplier $\Lambda(n)$ can be inferred from
the saddle-point equation at $\sum_a \sigma_a =0$. In the limit $n\to 0$
we find 
\begin{eqnarray}
\lim_{n \to 0}\frac{1+\Lambda(n)}{n}
=
\LL \log 2 \cosh ( \be h) \RR_h
 +\int \frac{\al\, \de J}{ 2 |J|^{\al+1}} \left [ \log \cosh( \be J)
 +\frac{1}{2} \LL \log (1- \tanh^2( \be J) \tanh^2( \be h)) \RR_h \right]
\end{eqnarray}
where the continuation to real temperatures has already been performed.
Using (\ref{eq:rsf}), (\ref{eq:seq_rep}), and (\ref{eq:RSrepl}) we
find back expression (\ref{eq:f_final}) for the free energy per spin. 
To obtain the internal energy we use  
\begin{eqnarray}
e&=& \lim _{N \to \infty} \frac{1}{N}  
 \overline{ \langle H( \{S_i \}) \rangle } =
 \lim_{ n \to 0}  \, \lim _{N \to \infty} 
 \frac{1}{N n} \sum_{\{S^{a}_i\}} 
  \frac{ \partial}{i \partial k} \; \overline{ \exp \left(
     ik \sum_{a=1}^{n}H(  \{ S_i^a\} )
   \right)}\; .
\end{eqnarray}
Again the limit $n \to 0 $ and the continuation to the real
temperatures may be performed and the result for the internal energy
obtained in section  \ref{sec:TDtrunc} gets reproduced.  

\subsection{The RS thermodynamic functions}

\begin{figure}
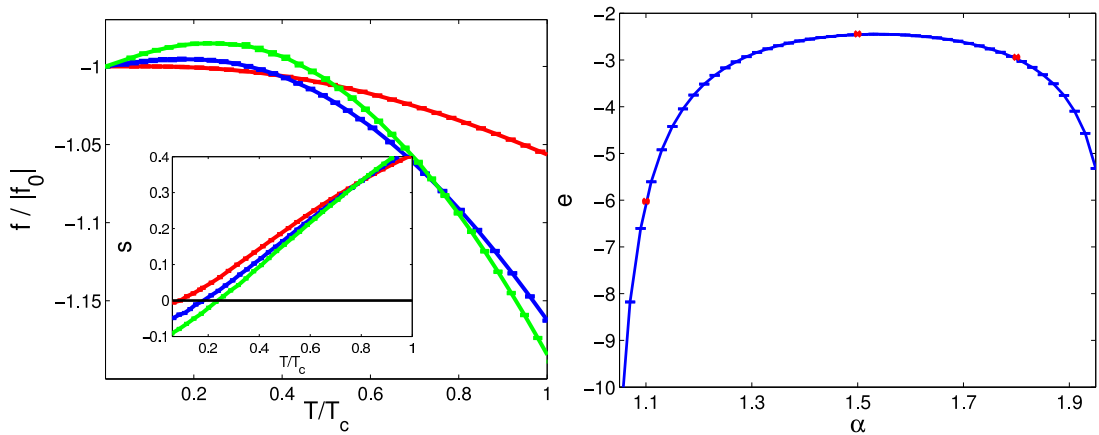

 \includegraphics[width=0.4\textwidth]{Freeenerg_Entrop.epsi}
 \includegraphics[width=0.4\textwidth]{GS.epsi}
 \caption{Thermodynamic functions of the L\'evy glass as calculated 
   from the numerical determination of the order parameters using
   population dynamics. Symbols give error bars of the statistical
   error intrinsic to population dynamics, lines are guide for the
   eye. Left: Replica symmetric free energy per spin for a L\'evy
   glass with $\al =1.8, 1.5, 1.1$ (green, blue, red). To lighten the 
   comparison the data have been normalized to the ground state energy 
   $f(T=0)$. The inset shows the corresponding results for the entropy
   per spin. As characteristic for spin glasses the replica symmetric
   entropy becomes negative at low temperature. 
   Right:  Replica symmetric ground-state energy per spin, $e=f(T=0)$,
   for a L\'evy spin glass as function of $\al$. The red symbols show
   the ground state energy for $\al =1.8, 1.5, 1.1$ within one-step
   RSB. As can be seen the corrections are rather small (cf. also 
   Fig.~\ref{Fig:GSE}).} 
  \label{Fig:fsg}
\end{figure}

In the previous subsections the thermodynamic functions of the L\'evy
spin glass within the assumption of replica symmetry were determined
using different approaches. In the left part of 
Fig.\ref{Fig:fsg} we have plotted the free energy per spin for zero
external field, $\hex=0$, as function of temperature for three
different values of $\al$. In the right part of this figure the
internal energy per spin in zero external field and at $T=0$ is shown
as function of $\alpha$. For small $T$ the numerically obtained values
for the free energy smoothly approach those for $e(T=0)$. The
RS free energy has negative slope for high temperatures, reaches a
maximum, and has a negative slope near $T=0$. Correspondingly the
entropy becomes negative at sufficiently low temperature, a well-known
signature for the breakdown of replica symmetry in spin glasses. We
have plotted the entropy per spin in the inset of 
Fig.~\ref{Fig:fsg}. Instead of using the derivative of $f(\be)$ with
respect to $T$ it is numerically much more accurate to determine
the entropy from the thermodynamic relation $s =\be(e-f)$. For
temperatures close to $T_c(\al)$ the curves for the entropy are
similar to those obtained within the Gaussian ansatz for the distribution
of local fields. At lower values of $T$, however, there are significant
deviations from the results obtained in \cite{CB}. In particular we
neither find a minimum of $s(T)$ at low temperatures nor do our data
extrapolate to $s(T=0)=0$. 

The temperature $T_{s=0}$ at which the entropy becomes negative
decreases with decreasing $\al$, cf.~Fig.~\ref{Fig:fsg}. This is
consistent with intuition since a larger fraction of strong bonds
should reduce the degree of frustration. We will find a similar
behavior when studying the influence of RSB in section
\ref{sec:1RSB}. Moreover we find for all $\alpha$ that $T_{s=0}$ is
much smaller than $T_c$. In \cite{CB} it was speculated that the
strong bonds in a L\'evy glass may stabilize a RS 
glass phase for some finite temperature interval below $T_c$. To
investigate this question we study the stability of RS in the next
section.  

\section{Stability of the RS Solution}\label{Sec:AT}

The self-consistency of the RS cavity approach can be tested by
investigating the correlations between spins. More precisely, the
divergence of the spin-glass susceptibility  
\begin{equation}\label{chisg}
\chi_{ SG} = \frac{1}{N} 
\sum_{(i,j)} \overline{  \Big(
 \langle S_i S_j \rangle 
-  \langle S_i \rangle \langle S_j \rangle 
 \Big)^2}
\end{equation}
signals the breakdown of replica symmetry \cite{BiYo}. 
In order to determine the stability boundary for the L\'evy spin glass 
we start again with the truncated model for which all weak bonds
smaller than $\eps$ are neglected. We then use techniques from the
theory of diluted spin glasses \cite{MezMonBook} to determine the
region of validity of RS and finally perform the \mbox{ $\eps \to 0$}
limit. As we will see this limit may be accomplished analytically
which makes the extrapolation back to the original model safe. 

\begin{figure}
  \begin{center}
   \includegraphics[width=0.45\textwidth]{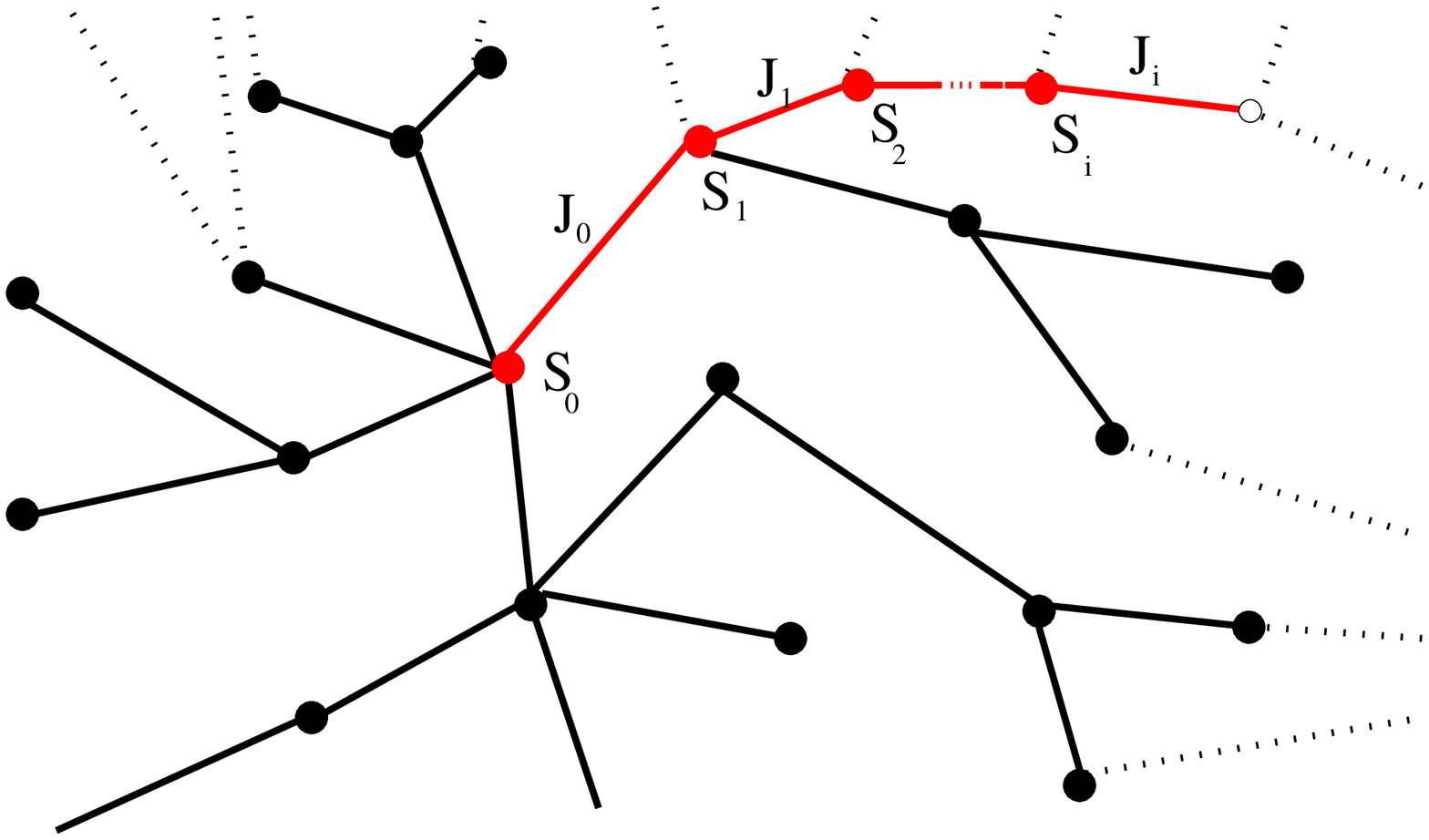}
   \hspace{+0.5cm}
    \includegraphics[width=0.45\textwidth]{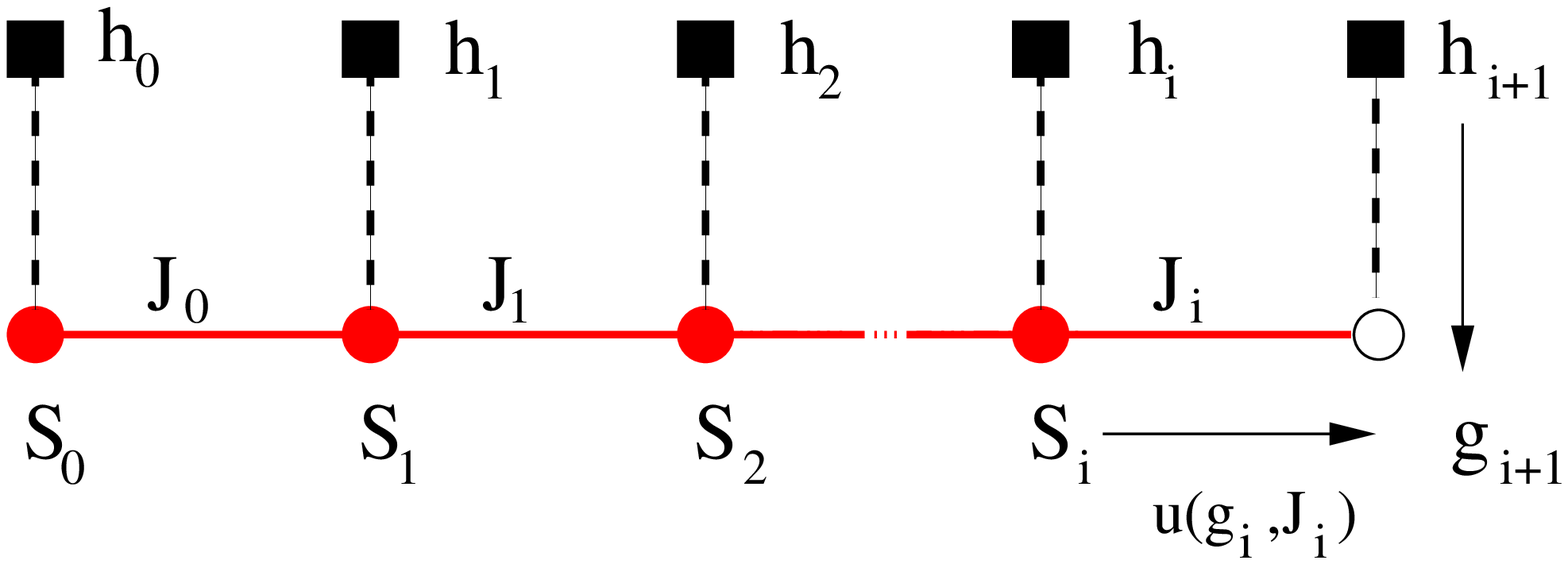}
 \end{center}
\caption{Left: Part of the graph corresponding to the truncated model.
       The couplings $J_i$ on the red  branch are treated exactly, 
       whereas the influence of all other couplings on the spins  
       of this chain is approximated by the local fields $h_i$ sampled
       from $\Ph(h)$. Right: The corresponding one-dimensional model
       defined by (\ref{eq:levy1d}).}
 \label{fig:AT1d}
\end{figure}

The underlying graph of the truncated model is locally a tree which
allows to write (\ref{chisg}) as 
\begin{equation}
\chi_{SG}=\sum_{r=1}^\infty \eps^{-\alpha r} \overline{C}_r\, ,
\label{eq:chiseries}
\end{equation} 
where  
\beq
C_r= \Big( \langle S_0 S_r \rangle -  
       \langle S_0 \rangle \langle S_r \rangle \Big)^2
\eeq
denotes the square of the connected correlation
function of two spins at a distance $r$ and $\eps^{-\alpha r}$ gives 
the average number of sites at distance $r$ from $i=0$. 
For large $r$ we expect $\overline{C}_r\sim \exp{(-r/\xi)}$ with some
correlation length $\xi$ and therefore the divergence of the
sum (\ref{eq:chiseries}) depends on whether
$\eps^{-\alpha}e^{-1/\xi}\lessgtr 1$. The stability analysis for the
fully connected L\'evy glass is thus mapped 
on the asymptotic behavior of the spin glass correlation
in a one-dimensional disordered Ising chain, cf.~Fig.~\ref{fig:AT1d}. It
is defined by the Hamiltonian \cite{MezMonBook}
\begin{equation}
E( \{ S_i\})=-\sum_{i=0}^{r-1} J_i S_i S_{i+1}-\sum_{i=0}^r h_i S_i  \,  ,
\label{eq:levy1d}
\end{equation}
where the couplings $J_i$ are distributed according to the
distribution $P_{\al,s}$ of strong bonds (\ref{eq:Pstrong}) and the external fields $h_i$
are drawn independently from the distribution $\Ph(h)$ determined in
section \ref{sec:SolSC}. 

The correlations in the one-dimensional system defined by
(\ref{eq:levy1d}) have been studied already in \cite{MonWei} using 
transfer matrix techniques for the replicated systems.  
We re-investigate the problem by using a cavity approach which leads to
the same result but sheds light on the importance of rare
fluctuations. Let us start again with an update equation of the form 
(\ref{eq:up}) for the local fields $\{g_i\}$ of model (\ref{eq:levy1d}).
\begin{equation}
g_{i+1}= h_{i+1}+ u(g_i,J_i)\,.
\label{eq:Up1}
\end{equation}
According to linear response theory the correlation function 
$C_i$ is related to the change of magnetization at site $i$ due to
a perturbation $ \delta h$ of the local field at site $0$. Denoting
by  $ \LL  \,\cdot \, \RR^\delta$  the canonical average induced by
the Hamiltonian 
\beq
  E_\delta( \{ S_i\})  =E( \{ S_i\}) - \delta h S_0
\eeq
we hence have 
\begin{equation}
\left(\left.
	\frac{ \partial \LL S _i \RR^{\delta}}{ \partial\,  \delta h }
	\right \vert_{\delta h = 0 }
	\right)^2
        =  \beta^2 [1-\tanh^2(\beta g_i)]^2  \left( \frac{
	\partial  g_i}{ \partial g_{0}} \right)^2=:  
        \beta^2 \cosh^{-4}(\beta g_i)\,   D_i 
\end{equation}
Using (\ref{eq:Up1}) and the chain rule we can derive a recursion
relation for $D_i$:  
\beq
 D_{i+1} =  \left( \frac{
  \partial u(  g_i,J_i)}{ \partial  g_{i}} \right)^2  D_i \, .
\label{eq:Up2} 
\end{equation} 
The quantities $g_i$ and $D_i$ are hence correlated random variables
due to their dependence on the quenched disorder represented by the
fields $h_i$ and the couplings $J_i$. From the update rules
(\ref{eq:Up1}) and (\ref{eq:Up2}) we find for their (site
dependent) probability distributions
\begin{eqnarray}  \nonumber 
P_{i+1} (g_{i+1}, D_{i+1}) 
&=&
\int \de J_i\, P_{\al,s }(J_i)
\int \de h_{i+1} \Ph (h_{i+1})
\int \de g_i
\int \de D_i \, P_{i} (g_{i}, D_{i})\\
&& \hspace{+2cm}
\delta \Big (g_{i+1}- [h_{i+1}+ u(g_i,J_i)] \Big)\,
\delta \Big(D_{i+1}-\left(\frac{ \partial u(g_i,J_i)}{ \partial
  g_i}\right)^2 D_i\Big).  
 \label{eq:Up3}
\end{eqnarray}
This equation describes how a perturbation propagates through the chain,
and thus contains information on how the correlation
$C_r=\cosh^{-4}(\beta g_r)\,D_r$ decays with
increasing $r$ in a given sample. It could easily be studied with
population dynamics in order to get the typical decay rate. However,
in order to characterize the spin-glass susceptibility
(\ref{eq:chiseries}), we need the behavior of the {\it average} of
$C_r$, not the typical one, and this average correlation is dominated
by rare instances of the couplings. 

In order to obtain the behavior of the average correlation we first
derive a recursion relation for the auxiliary quantity 
\beq
I_i(g_{i}) := \int \de  D_i \,  P_{i} (g_{i}, D_i)\, D_i 
\eeq
from which the averaged correlation function can be obtained by
integration, $\overline{C_r}  = \int  \de g\, \cosh^{-4}(\beta
g)\, I_r(g)$. The average decay of correlations is hence determined by
the $i$-dependence of $I_i(g)$. We now find from (\ref{eq:Up3})
\begin{eqnarray} \nn
I_{i+1}( g_{i+1} ) 
 &=& \int \de J_i \, P_{\al, s }(J_i)
 \int \de h_{i+1} \Ph (h_{i+1})
 \int \de g_i
 \int \de D_i\, P_{i} (g_{i}, D_{i})
  \left(\frac{\partial u(g_i,J_i)}{ \partial g_i}\right)^2 D_i\;
 \delta\left( g_{i+1}-[h_{i+1}+u(g_i,J_i)]\right) \\ \nn
 &=&\int \de g_i \int \de J \, P_{\al, s }(J) \int \de h\, \Ph (h) 
 \left(\frac{\partial u(g_i,J)}{ \partial g_i}\right)^2 
\delta\left( g_{i+1}-[h+u(g_i,J)]\right)\; I_i(g_i)  \\ 
&=:&\int \de g_i \,  K( g_{i+1}, g_{i})  \, I_i(g_i)   \, .
 \label{transfer}
\end{eqnarray}
from which we infer that the asymptotic behavior of $I_i(g)$ and hence
also that of the averaged correlation function $\overline{C}_r$ for
$r\to\infty$ is characterized by the largest eigenvalue $\nu$ of the
transfer matrix $K(x,y)$ defined in the last step of (\ref{transfer}).  
We therefore conclude that the stability of RS is determined by the
convergence of the geometric series 
\begin{equation}
\chi_{SG} = \sum_{r=1}^\infty{ \Big( \eps^{-\alpha} \nu \Big)
}^{r} =:  \sum_{r=1}^\infty\lambda ^{r} \, .
\label{chiseries2}
\end{equation} 
The most convenient way to determine the stability parameter $\lambda$
is to calculate the largest right eigenvalue of $\eps^{-\al}K$, i.e. to
solve the equation 
\beq 
\lambda \, \phi( x ) = 
      \eps^{-\al} \int \de y \, K^T(x,y) \, \phi( y ) =\int \frac{
      \al\, \de J }{2 |J| ^{\al+1}}  
 \int \de h \, \Ph (h) \left(\frac{\partial u(x,J)}{ \partial
   x}\right)^2 \phi (h+u(x,J)  ) \, .
\label{eq:eigenvalue}
\eeq
where $K^T$ denotes the transposed operator, and the $\eps\to 0$ limit
has been taken in the last expression (notice that it can be taken
safely 
since the function $(\partial u(x,J)/\partial x)^2$ behaves as $J^2$ 
for small $|J|$). 
For $\lambda\leq 1$ RS is stable, otherwise it is unstable. The
correlations in the one-dimensional system defined by
(\ref{eq:levy1d}) have been studied already in \cite{MonWei} using
transfer-matrix techniques for the replicated systems. The eigenvalue 
problem for the replicon mode considered there coincides with
(\ref{eq:eigenvalue}). 

The determination of the largest eigenvalue in (\ref{eq:eigenvalue})
has to be done numerically and can be accomplished by straight
iteration. Starting with an arbitrary positive function $\phi^{(0)}$
we use  
\begin{eqnarray} \label{eq:lait}
\phi^{(m)}( x )
&=&
\frac{1}{Z_m}\int \frac{ \al \, \de J }{2  |J| ^{\al+1}} 
 \int \de h \, \Ph (h)
 \left(\frac{\partial u(x,J)}{\partial x}\right)^2
\phi^{(m-1)}(h+u(x,J)  ),
\end{eqnarray}
and impose the normalization  $\int \phi^{(m)} (x) \, \de x =1 $ after
each step. After many iterations $\phi^{(m)}$ converges to the eigenvector of 
the operator (\ref{eq:eigenvalue})  with the largest eigenvalue which
in turn is given by $\lambda  = \lim_{m\to  \infty} Z_m $. The
complete numerical procedure is hence as follows. For given $\beta=1/T$
and $\hex$ we calculate the distribution of local fields $\Ph(h)$ from
(\ref{eq:sc-eq}) as described in section \ref{sec:SolSC}. We use
$\Ph(h)$ to determine the transfer matrix $K(x,y)$ according to
(\ref{transfer}) and extract its largest eigenvalue
$\lambda(T,\hex)$ from the iteration (\ref{eq:lait}). The AT-line
is then implicitly given by $\lambda(\hex,T_{AT})=1$. 

\begin{figure}
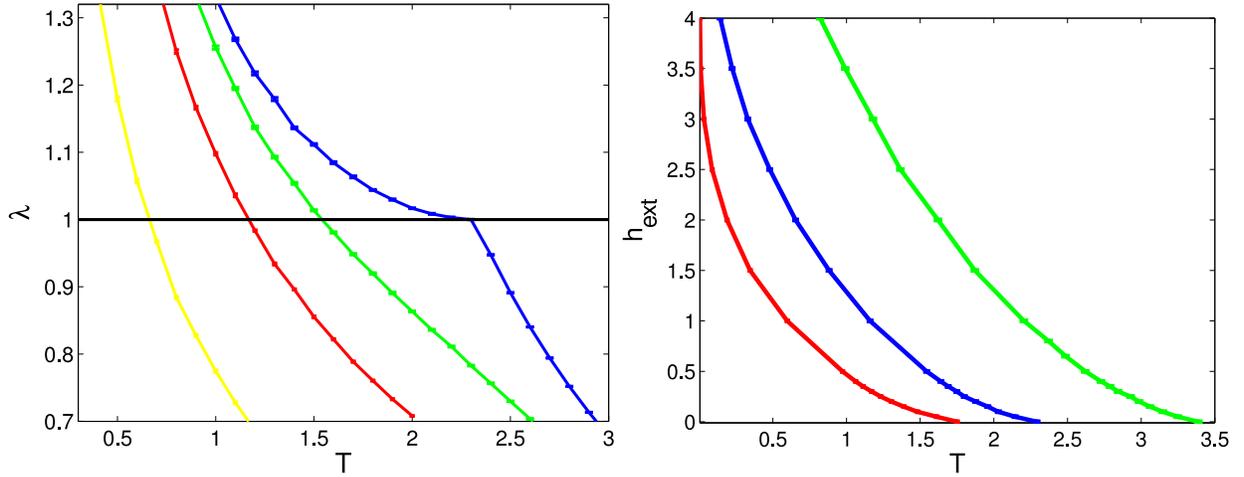

  \begin{center}
    \includegraphics[width=0.45\textwidth]{lambda_T_15.epsi}
    \includegraphics[width=0.45\textwidth]{AT_Tabs.epsi}
 \end{center}
\caption{ Determination of the de Almeida-Thouless line. Left:
  Stability parameter $\lambda$ as function of temperature for
  $\al=1.5$ and $\hex=0,\, 0.5,\, 1$ and $2$ (from right to
  left). From the intersection of the curves with the stability
  boundary $\lambda=1$ the AT-line is determined.
  Right: Phase diagram  of a L\'evy spin glass with 
  $\alpha=1.8, \, 1.5,  \,1.1 $ (from right to left).
  Above the AT-lines shown RS is stable,
  below it is unstable.}
  \label{fig:AT}
\end{figure}

In the left part of Fig.~\ref{fig:AT} we have plotted the stability
parameter $\lambda$ as function of $T=1/\be$ for different values of
$\hex$. One finds that $\lambda(T,\hex)$ is, for all values of $\hex$, a
monotonically decreasing function of temperature and crosses the line
$\lambda=1$ at exactly one point. The collection of these points
defines the AT-line $T_{AT}(\hex)$ shown in the right part of
Fig.~\ref{fig:AT} for three values of $\al$.  

Of particular interest is the value of $T_{AT}$ in zero external
field. In the paramagnetic region characterized by $\Ph(h)=\delta(h)$
the stability parameter may be determined analytically. Using
$u(0,J)=0$ we find in this case from (\ref{eq:eigenvalue}) for $x=0$  
\begin{equation} 
\lambda \phi ( 0 )=
\left.\int_0^{\infty} \frac{ \al\, \de J }{ J ^{\al+1}} 
\int \de h \, \delta (h)
\left(\frac{\partial u(x,J)}{\de x}\right)^2\right |_{x=0}
\phi (h+u(0,J))
=
\int_0^{\infty} \frac{ \al\, \de J }{ J ^{\al+1}} \tanh^2(\be J)\phi(0)
=
\left(\frac{T_c(\al)}{T}\right)^{\al}\phi(0),
\label{eq:phi0}
\end{equation}
Assuming $\phi(0)\neq 0$ we hence find the instability of the
paramagnetic solution at $T_c(\al)$. Moreover, our numerical results
indicate that $\lambda(T,0)>1$ for all $T<T_c(\al)$.  

Because the numerics is somewhat subtle when $\Ph(h)$ is near to a
$\delta$-function we corroborate this result by a perturbative study
of the eigenvalue problem to leading order in the reduced temperature
$\tau: = 1- T/T_c(\al)$. To this end we expand (\ref{eq:eigenvalue})
in $x$ up to order $x^4$. Using the symmetry of $\Ph$ and $\phi$ the
truncated eigenvalue equation acquires the form $\lambda \vec \phi =
\mathbf K \vec \phi $ with the vector $\vec{\phi} =
(\phi(0),\phi''(0), \phi''''(0))^T$. The matrix elements $K_{ij}$
depend on the moments of $\Ph$ which to the required order in the
reduced temperature read   
\begin{align} \nonumber
 \LL h^2\RR_h &= 
 \frac{\al}{2 }\, T_c^2(\al) \,\tau
 +
\frac{\al}{24} \, T_c^2(\al)\, \frac{\al (11\, t_{2,\al} +28\,
 t_{4,\al})  -18\,(t_{2,\al}- t_{4,\al}) } 
{  t_{2,\al}-t_{4,\al} } \,  \tau^2\\
\LL h^4\RR_h &= \frac{ 3 \,  t_{2,\al}} { t_{2,\al}- t_{4,\al} }  
   \left(\frac{\al}{2 } \right)^2 T_c^4(\al) \, \tau ^2 \, , 
\end{align}  
where we defined 
\beq\label{eq:deft}
t_{m,\al}:=\int_0^{\infty} \frac{ \al\, \de J }{ J ^{\al+1}}
\tanh^m(J) \, .
\eeq
For the largest eigenvalue of ${\mathbf K}$ we find
\beq 
 \lambda=1+
  \frac{t_{2,\al}+2\,t_{4,\al}}{t_{2,\al}-t_{4,\al}} \tau^2 
  +{\cal  O}(\tau^3) 
\eeq
The coefficient of the quadratic term is always positive, since 
$t_{m, \al}$ decreases with $m$ as implied by (\ref{eq:deft}) and 
$\tanh(x) \leq 1$. 

We therefore find $\lambda(T,\hex=0)>1$ for all $T<T_c(\al)$ in the
L\'evy spin glass. Correspondingly there is no stable replica
symmetric glass phase as proposed in \cite{CB,Cizeau}. The AT-line for 
the L\'evy spin glass as shown in the right part of Fig.~\ref{fig:AT}
looks indeed qualitatively similar to other spin-glass models. 

We finally comment on the assumption $\phi(0)\neq 0$ made after
eq.~(\ref{eq:phi0}).
For $\phi(0)=0$ we differentiate (\ref{eq:eigenvalue})
$m$ times where $\partial^m \phi$ is the first derivative with
$\partial^m \phi (0 ) \neq 0$.
Evaluating this equation at $x=0$ in the paramagnetic region i.e.
for $\Ph(h) =\delta(h)$ we infer that the corresponding eigenvalue
$\lambda_m$ is given by 
\beq  \nonumber
\lambda_m=\int_0^{\infty} \frac{ \al \,  \de J }{ J ^{\al+1}}
\tanh^{2+m}(\be J)=\frac{t_{(2+m),\al} }{T^\al} \; .
\eeq
Since $t_{m, \al}$ decreases with $m$, it results that eigenfunctions with $\phi(0)=0$
give rise to eigenvalues which are smaller than those for
eigenfunctions with $\phi(0)\neq 0$ and are hence not relevant for
the stability problem at hand. 

\section{One-step RSB}\label{sec:1RSB}

The RS solution fails at low temperatures as it ignores the
possibility of several pure states 
\cite{MezParBethe}.  
We therefore consider a solution which takes into account the
existence of many pure states and corresponds to the one-step RSB
solution in the replica formalism. The local fields $h_i^{\gamma}$
corresponding to the different states $\gamma$ are assumed to be
independent random variables sampled from a site dependent
distribution $P_i$. After averaging over the disorder, the natural 
order parameter is the probability distribution $\Q$ of the local 
field distributions $P_i$, and one can derive a self-consistent
equation for it. In this section we work at zero temperature, where
RSB effects should be  most pronounced, and we show how to compute the
ground-state energy density of the model within the one-step RSB. For
simplicity we shall keep to the case of zero external field. The
method which we use  to derive expressions for the ground-state energy
consists in using the truncated model, to which the general one-step
RSB approach to diluted models developed in \cite{MMGP} can be
applied. The limit $\eps \to 0$ is performed analytically at the end
of the calculations. We shall derive the complete one-step RSB
equations and solve them numerically using an approximation called the
factorized approximation \cite{WongSherr}. 

The assumption of one-step RSB is the following: for a fixed
realization of the disorder the local fields $h^{\gamma}_i$ on a given
site $i$
are random variables due to the existence of many local ground
states. We denote by $P_i$ the corresponding site dependent
probability distribution. 

When iterating, i.e. merging $K$ spins at a new site, the update rule 
for the probability distribution at this new site reads  
\begin{eqnarray}
P_{new}(h_{new})
=
C \int \prod_{i=1}^{K} \de h_i \, P_{i}(h_i) \,
 \delta \Big( h_{new}- \sum_{i=1}^K  u_0(J_i,h_i) \Big) \,
 \exp\Big( \mu \sum_{i=1}^K     
\max(|J_i|-|h_i|,0)
  + \mu |h_0|\Big),
\label{eq:up6}
\end{eqnarray}  
where $\mu$ is the one-step RSB parameter, the constant $C$ ensures
the normalization of $P_{new}$, and $u_0(h,J)$ is the zero temperature
limit  of (\ref{eq:up}) \mbox{$u_0(h,J)= \lim_{\be \to \infty }
  u(J,h)=\min(|h|,|J|) \, $}.  
(Notice that  we use a slightly different notation from the one in \cite{MMGP}:
we consider the update rules for the
probability distributions \mbox{ $\widetilde P_i (h_i) =  c_i P_i(h_i)
\exp(-\mu |h_i|)$}, instead of $ P_i(h_i)$ used in \cite{MMGP}; this
is to ensure a safe $\eps\to 0$ limit).

The ground-state energy 
of the L\'evy spin glass is obtained within the one-step RSB
approximation by maximizing
\begin{eqnarray} 
\Phi( \mu )=  \lim _{\eps \to 0 }\Phi_\eps( \mu ) = \lim _{\eps \to 0 } \left [
 \overline{\dE E^{site}_\eps(\mu)}
 -\frac{1}{2}\overline{ (K+1)\dE E^{bond}_\eps(\mu)} \right]
\label{eq:Phi_eps}
\end{eqnarray}
with respect to $\mu$, where the energy shifts $ \dE
E^{site}_\eps(\mu)$ and $\dE E^{bond}_\eps(\mu) $  are given below.

The energy shift corresponding to a site addition reads 
\begin{eqnarray}
 \dE E^{site}_\eps (\{P_i,J_i\}| \mu)&=&- \frac{1}{\mu} \log
\int \prod_{i=1}^{K+1} \de h_i \,  P_i(h_i)  \, 
\exp \left  (
             \mu \sum_{i=1}^{K+1} \max(|J_i|-|h_i|,0)+
             \mu  \left \vert \sum_{i=1}^{K+1} u_0(J_i,h_i)\right \vert
      \right) \\ \nn
&=& \frac{1}{\mu}\log C = \log \int \de H_0 \,  P_0(H_0) \,  e^{-\mu
	|H_0|  } - \sum_{i=1}^{K+1} \log  \int \de h_i \,  P_i( h_i)
      \, \exp(\mu \max(|J_i|-|h_i|,0))\, , 
\end{eqnarray}
where  $ P_{0}$ denotes the distribution of the fields $H_0 =
\sum_{i=1}^{K+1}u_0(J_i,h_i)  $  acting on the new spin connected to
$(K+1)$ old ones.   

The energy shift due to a bond addition amounts to 
\begin{eqnarray} \nn
\dE E^{bond}_\eps (P_1,P_2,J| \mu)
&=&
- \frac{1}{\mu} \log \int  \de h \,\de h'  \, P_1(h) \, P_2(h') 
\exp \left (
            \mu  \max_{\sigma \sigma'} \big(h \sigma +h' \sigma' + J
	    \sigma \sigma' \big)  -\mu| h|-  \mu |h'|
     \right) \\
&=& - |J| -  \frac{1}{\mu}
      \log  \int \de h \, \de h'  \, P_1(h) \, P_2(h') \,
      \Big[ 1+ \theta(-J h h') \big(e^{-2 \mu \min(|J|,|h|,|h'|) } -1
	\big) \Big] \,. 
\end{eqnarray} 
After the average over disorder, the added site has the same properties
as the old ones, and a self-consistency equation for the order parameter
$\Q$ can be  derived: in (\ref{eq:up6}) $P_{new}$ must have the
same distribution  $\Q$ as the  $ \{ P_i \}$.
Assuming that $P_{0}$ is also  from the same distribution $\Q$ (which
is the case in the  $\eps \to 0 $  limit) the averaged energy shifts
read 
\begin{eqnarray} 
 \overline{\dE E^{site}_\eps (\mu) } 
&=& \frac{1}{\mu}
\LL
      \log \int \de h \,  P(h)\, e^{-\mu |h|}
      -\eps ^{-\al} \int P_{\al , s }(J)\,
      \log \left(
      1+ \int \de h \, P(h) \, \big [ e^{ \mu \max(|J|-|h|,0)} -1  \big]
      \right)
\RR_P 
\end{eqnarray}
and
\begin{eqnarray} 
\overline{ 
\dE E^{bond}_\eps (\mu)} 
 &=&
 \int \de J \,  P_{\al,s}(J) 
\LL
|J| +\frac{1}{\mu} \log  \int \de h \, \de h'  \, P(h)  \,P'(h') 
      \Big[ 1+ \theta(-J h h') \big(e^{-2 \mu \min(|J|,|h|,|h'|) } -1
	\big) \Big] \RR_{P,P'}
\label{eq:E2}
\end{eqnarray}
respectively, where $\LL \, \cdot \, \RR_P$ denotes an average in
which $P$ is drawn from the probability distribution $\Q$.

We have now all ingredients entering $ \Phi_\eps(\mu)$ of the
truncated model and the  limit $ \eps \to 0$ can be performed to
obtain the corresponding expression for the L\'evy spin glass. The
symmetry of the averaged field distributions is crucial in this case
as  
\begin{eqnarray} 
\LL
 \int \de h \,\de h' \, P(h) P'(h') \theta(- h h')
\RR_{P,P'}
=\frac{1}{2} 
\end{eqnarray}
leads to a  convergent integral in (\ref{eq:E2}) at small values of $J$.
One obtains 
\begin{eqnarray}  \nn
 \Phi( \mu ) 
&=&
\frac{1}{\mu}
\LL  \log  \int  \de h \, P(h) \, e^{-\mu|h|} - \int_0^{\infty} \frac{
  \al \, \de J }{ J^{\al+1}}   \log \left(
          1+ \int_{-J}^J\de h \, P(h)\big [e^{\mu(|J|-|h|)}-1\big ]
       \right) \RR_P
\\  \nn
&&
+\int_0^\infty  \frac{\al \, \de J}{ J^{\al+1}}
\LL
\frac{J}{2}  +\frac{1}{2 \mu} \log  \int \de h \, \de h' \, P(h) P'(h') 
      \Big[ 1+ \theta(- h h') \big(e^{-2 \mu \min(|J|,|h|,|h'|) } -1
	\big) \Big] \RR_{P,P'} \, .
\end{eqnarray}

Notice that the self-consistent equation for the mean $P$ with respect
to  $\Q$, denoted by $\bar P (h)=\int dP\, \Q[P]\, P(h)$, coincides, in the
limits  $ \eps \to 0$ and $\mu \to 0 $, with the self-consistent
equation (\ref{eq:sc-eq}) for the RS order parameter $\Ph$ for zero
temperature and zero external field. In particular one recovers in
this limit: 

\begin{equation}  
 \lim_{\mu \to 0} \Phi( \mu )=
 \lim_{\be \to \infty } f_{RS}(\be, \hex = 0)  
\end{equation}

\subsection{Factorized approximation}
\begin{figure}
\begin{center}
 \includegraphics[width=0.48\textwidth]{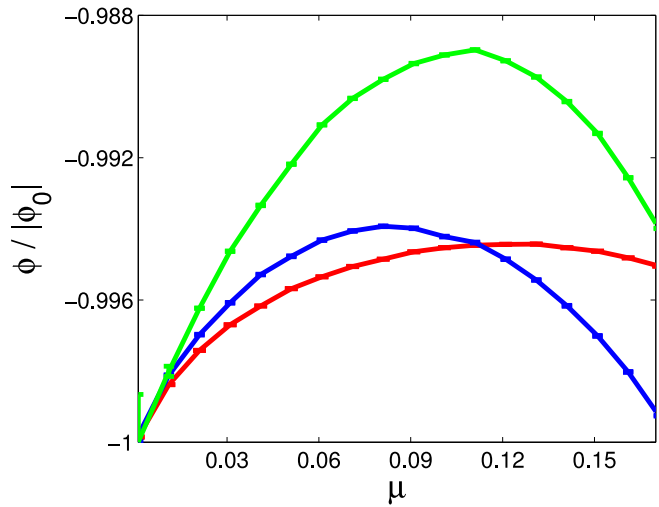}
\end{center}
\caption{Trial ground energy $\Phi_f(\mu)$ of a L\'evy glass as
        function of the one-step RSB parameter $\mu$ for  $\al = 1.1 $
        (red), $ 1.5$ (blue), and $1.8$ (green) as obtained within the
        factorized approximation. Symbols show population dynamics
        results with error bars, lines are guides for the eye. The
        curves are normalized to the modulus of the RS ground state
        energy $|\Phi_f(\mu=0)| $. The ground state energy in one-step
        RSB is given by the maxima of the curves which is about one
        per cent higher than the RS result.} 
 \label{Fig:GSE}
\end{figure}

It is numerically rather heavy to sample $P$ from the distribution
$\Q[P]$. The factorized approximation makes the task much easier by
confining the space of all distributions to one distribution  $\bar
P$, i.e. it assumes that $\Q$ is a functional delta-function. Within
this ansatz a population dynamics algorithm can be easily applied to
determine $ \bar P$ for all values of the parameter $\mu$.  
One then has to find the maximum of the function 
\begin{eqnarray}
 \Phi_{f}(\mu)&=& \frac{1}{\mu} \log \int \de h \,\bar{ P}(h) e^{-\mu
 |h|} +\int_0^\infty \frac{\al\, \de J}{J^{1+\al}} R_{\mu, \bar P}(J) 
\label{eq:Phi_fac}
\end{eqnarray} 
where  $R_{\mu, \bar P }$ denotes 
\begin{eqnarray} \nn 
 R_{\mu, \bar P}(J)&=&
 -  \frac{1}{\mu}  \log \Big( 1+ \int_{-J}^{J} \hs{-0.2cm} \de h \,
 \bar  P(h)     \big[  e^{\mu(J-|h|) }-1   \big]
\Big) + \frac{J}{2}  \\ \nn &&
+ 
\frac{1}{ 2 \mu  }
\log \int   \de h \, \de h' \, \bar  P(h)  \,  \bar P(h') \,
   \Big[
     1+ \theta(- h h') 
      \big(
       e^{-2 \mu \,  \min(J,|h|,|h'|)}-1
      \big)
   \Big] .
\label{eq:R_mu}
\end{eqnarray} 
An expansion of $ R_{\mu, \bar P }$ at small  and large values of $J$
allows to perform the $J$ integral in (\ref{eq:Phi_fac}) analytically
in the corresponding regions. For small values of $J$ one obtains 
$ R_{\mu , \bar P}(J) \approx  \frac{1}{4} \mu J^2$, whereas in the
large $J$ limit one has 
\begin{eqnarray} \nn 
 R_{\mu, \bar P }(J) & \approx& 
-\frac{J}{2} 
-\frac{1}{\mu} \log  \int \de h \,  \bar P (h) \,  e^{-\mu |h|}
+\frac{1}{2\mu} \log \left( 1+ \frac{1}{2} \int \de h \, \de h' \,
\bar P (h)  \,\bar P(h')   
 \big[
 e^{-2 \mu  \, \min (|h|,|h'|)}-1
 \big]\right) \,.
 \end{eqnarray} 
A numerical test of the result for large $J$ shows that the asymptotic
behavior is reached for $J>50$. Using this decomposition, the
computation of $\Phi_f(\mu)$ becomes numerically easy. We have
obtained the distribution $\bar P$  by a   population dynamics
algorithm after performing $10^4$ iterations for a population of
$10^5$ fields. The result is shown in  Fig.~\ref{Fig:GSE}, which 
plots the relative increase  $\frac{\Phi_f(\mu)}{ |\Phi_f(0)|}$  
compared to the RS solution as a function of the one-step RSB
parameter $\mu $. 
The maximal value of $\Phi_f(\mu)$ gives the estimate for the ground
state energy of the L\'evy spin glass within the one-step RSB
factorized approximation. We see that the relative corrections to this
ground state energy due to
RSB remain rather small, of the order of one to two per cent, which is
comparable to the typical corrections found for instance in the
SK model \cite{SKcorr}. It is also smaller when
$\alpha$ is close to $1$, which agrees with the intuition according to
which a smaller value of $\alpha$ leads to a stronger hierarchy of
couplings and therefore to a lesser degree of frustration.
\section{Conclusion}
We have presented a detailed study of the properties of the L\'evy spin
glass at the replica symmetric level at all temperatures and magnetic
fields, and a one-step RSB study at zero temperature and zero external
field. One main ingredient of this
study has been the introduction of a truncated model where the
couplings with values smaller than a cutoff $\eps$ are neglected. The
truncated model naturally enters the category of dilute spin glasses
for which  various techniques have been developed in recent years,
allowing for a detailed analysis. The $\eps\to 0$ limit requires some
care, and complicates notably the analysis with respect to the studies
of 'usual' spin glasses, but we have shown that it can be controlled.

The physical picture which has been obtained shows a
spin glass behavior which is generally much closer to the standard
behavior found in the SK or in diluted models
than what had been claimed before. Within the RS approximation,
the entropy decreases with temperature and becomes negative at low
temperatures, but does not turn back to $0$ when $T\to 0$. The AT
instability line can be computed, and the whole spin glass phase turns
out to be unstable with respect to RSB effects. On the other hand, the
quantitative effects of RSB on the ground state energy are relatively
small and become smaller with decreasing the L\'evy exponent $\alpha$.

\acknowledgments
We would like to thank Jean-Philippe Bouchaud, Helmut Katzgraber, and
Martin Weigt for interesting discussions. Financial support from the
Deutsche Forschungsgemeinschaft under EN 278/7 is gratefully
acknowledged. MM thanks the Alexander von Humboldt foundation for its
support.

\end{document}